\documentclass[10pt]{iopart} 

\usepackage{amsfonts} 
\usepackage{graphicx} 

\usepackage{hyperref} 
\hypersetup{
    unicode=true,                                           
    a4paper=true,
    plainpages=false,
    pdffitwindow=true,                                      
    pdftitle={},                      
    pdfauthor={},                            
    pdfsubject={},                                          
    pdfkeywords={},
    colorlinks=true,                                        
    linkcolor=blue,                                         
    citecolor=blue,                                         
    filecolor=blue,                                         
    urlcolor=blue                                           
}
\newcommand{\urlNewWindow}[1]{\href[pdfnewwindow=true]{#1}{\nolinkurl{#1}}}

\begin{document}

\bibliographystyle{apsrev4-1}

\title{Measuring the Faraday effect in olive oil using permanent magnets and Malus' law}

\author{Daniel L Carr, Nicholas L R Spong, Ifan G Hughes and Charles S Adams}

\address{Joint Quantum Centre (JQC) Durham--Newcastle, Department of Physics,  Durham University, South Road, Durham, DH1 3LE, United Kingdom}
\date{\today}

\ead{daniel.l.carr@durham.ac.uk}

\begin{abstract}
We present a simple permanent magnet set-up that can be used to measure the Faraday effect in gases, liquids and solids. By fitting the transmission curve as a function of polarizer angle (Malus' law) we average over fluctuations in the laser intensity and can extract phase shifts as small as $\pm$ 50 $\mu$rads. We have focused on measuring the Faraday effect in olive oil and find a Verdet coefficient of \textit{V} = 192 $\pm$ 1 deg T$^{-1}$ m$^{-1}$ at approximately 20 $^{\circ}$C for a wavelength of 659.2~nm. We show that the Verdet coefficient can be fit with a Drude-like dispersion law $A/(\lambda^2 - \lambda_0^2)$ with coefficients $A$ = 7.9 $\pm$ 0.2 $\times$ 10$^{7}$ deg T$^{-1}$ m$^{-1}$ nm$^2$ and $\lambda_0$~=~142~$\pm$~13~nm.
\end{abstract}

\noindent{\it Keywords\/}: Faraday effect, Verdet coefficient, Malus' law, olive oil

\section{Introduction}

The Faraday effect (see e.g. \cite{f2f}) has a wide range of applications including filtering \cite{Ohman,Dick} and limiting feedback in optical systems \cite{Weller}; it also affects light propagating through interstellar media \cite{Gardner}. The effect occurs when a magnetic field parallel to the direction of light propagation induces a circular birefringence in a medium which causes linearly polarized light to rotate as it travels through the medium. The magnitude of this rotation ($\theta$) may be expressed as $\theta = VBl$, where $B$ is the magnetic field strength, $l$ the path length and $V$ is a material-dependent factor known as the Verdet coefficient \cite{f2f}.

Studying the Faraday effect (see e.g. \cite{mit,rutgers}) and other examples of circular birefringence such as optical rotation in sugar solutions  \cite{Compton,Mahurin,Nixon} is a common experiment in undergraduate teaching labs, as it elegantly links the concepts of polarization, dichroism, birefrigence and scattering \cite{f2f}. However, measuring the Faraday effect can be challenging, especially in cases where the Verdet coefficient is small or the optical path length is short. There are many different methods proposed in the literature, some utilizing DC fields \cite{Loeffler,Pedrotti,Hunte,Jacob,VanBaak} and some AC fields \cite{Jain,Valev,Valev2,Phelps,Chang,Brandon, Duffy}; by modulating the field and using a lock-in amplifier it is possible to extract rotation angles in the range of microradians. Most experiments employ electromagnets producing fields typically in the range of up to 0.025 T \cite{Jain,Brandon} although experiments with larger fields have been performed \cite{Pedrotti}. 

In this paper, we report on a simple Faraday measurement apparatus based on small permanent magnets where fields over 0.6 T are easily accessible. The experiment is inexpensive as it does not require the use of an electromagnet or lock-in amplifier. We show that by fitting a Malus' law transmission curve it is possible to measure polarization rotations as small as 50 $\mu$rads without the need for field modulation. Consequently, the DC Faraday effect is observable even in weakly magneto-optical materials.

To demonstrate the capabilities of our apparatus, we have measured the Verdet coefficient of olive oil, which was of interest due to a suggestion in the literature of an anomalously high Verdet coefficient at 650 nm \cite{Shakir}, contradicting more recent measurements \cite{Brandon}. Our measurements show that Verdet coefficient of olive oil is similar to other liquids such as water and much smaller (about 2 rad T$^{-1}$ m$^{-1}$ at 780~nm) than Faraday active magneto-optical materials such as the crystal TGG (82~rad~T$^{-1}$~m$^{-1}$) and resonant media such as Rb vapour (1.4$\times10^3$~rad T$^{-1}$~m$^{-1}$) \cite{Weller}.

\section{Theory}
From Malus' law (see \cite{f2f}) we may describe the intensity ($I$) of a beam passing through two crossed polarizers with \begin{equation}
    I = I_0 \cos^2{\phi},
\label{Malus}
\end{equation}
where $I_0$ is the peak intensity and $\phi$ is the angle between the two polarizers. Light passing through a circularly birefringent medium has its plane of polarization rotated by an angle $\theta$. When the optically active medium is placed between the polarizers, we may describe the light transmitted through the polarizers using \begin{equation}
    I = I_0 \cos^2(\phi + \theta) + c.
\label{ModifiedMalus}
\end{equation}
Here we have introduced an offset $c$ to account for imperfect extinction of the polarizers and the presence of background light.

The magnitude of rotation for a given magnetic field strength ($B$) and path length (d$l$) is material-dependent and classified by the Verdet coefficient ($V$) such that \begin{equation}
    \theta = V \int_0^l B \mathrm{d} l = V B_{\rm av} l,
\label{Verdet}
\end{equation} 
where $B_{\rm av}$ represents the average magnetic field strength across the sample and $l$ is the sample's total length. Due to its relation to the material's refractive index \cite{VanBaak}, the Verdet coefficient is found to depend on the temperature and wavelength so may be modelled using a dispersion law. We tested this by determining the Verdet coefficient at multiple wavelengths and attempting to fit a dispersion curve. We compared the data to two separate models, the first being a Cauchy-type \cite{Cauchy1,Cauchy2} dispersion of the form \begin{equation}
    V(\lambda) = A + \frac{B}{\lambda^2},
\label{Cauchy}
\end{equation}
where $A$ and $B$ are both fitting parameters and $\lambda$ is the wavelength of light. Secondly we attempted to fit a Drude-type \cite{Drude} dispersion of the form: \begin{equation}
    V (\lambda) = \frac{A}{\lambda^2 - \lambda_0^2},
\label{Drude}
\end{equation}
where now $A$ and $\lambda_0$ are the two fitting parameters. We believed using the Drude model would provide a more profound insight into the underlying physics as the value of $\lambda_0$ should correspond to an absorbance peak.

\section{Methods} 
\label{Methods}

\begin{figure}[t]
\begin{center}
\includegraphics[trim = 0mm 0mm 0mm 0mm, clip, width=12cm]{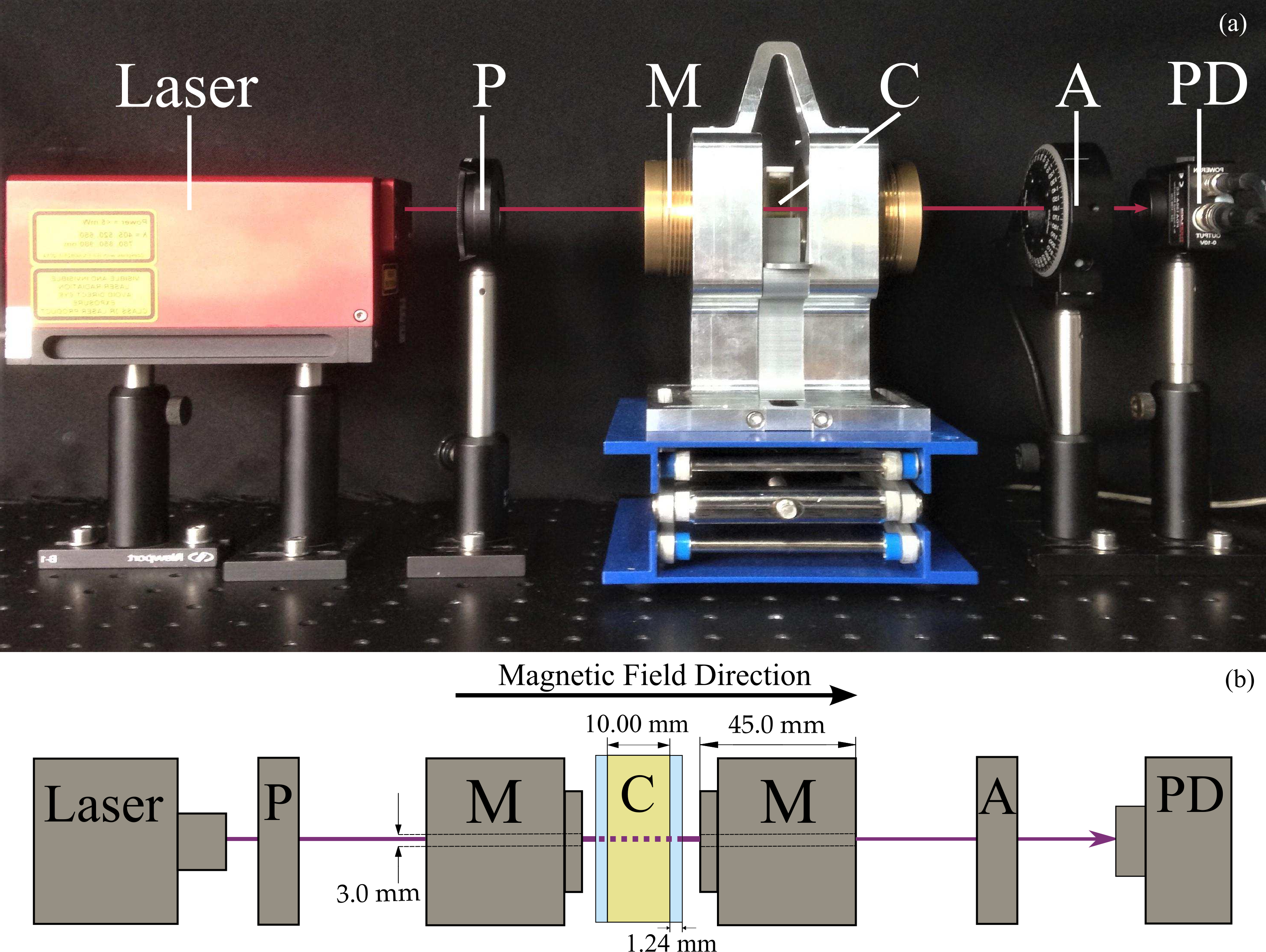}
\caption{
The arrangement of the apparatus. (a) A photograph of the equipment aligned on an optical breadboard. The arrow shows the direction of laser propagation. P =  polarizer, M = permanent magnet, C = glass cuvette, A = analyzer, PD = photodiode. The analyzer and photodiode were both connected to a PC, which could automatically rotate the analyser and read voltages from the photodiode. (b) A schematic drawing displaying the dimensions of the annular magnets and the cuvette.
}
\label{fig:1}
\end{center}
\end{figure}

The experimental apparatus consisted of simple optical equipment, commonly available in an undergraduate laboratory. The arrangement of the equipment is shown in \fref{fig:1}. Measurements of optical rotation as a function of the magnetic field strength were obtained at 7 different wavelengths (405.4 nm, 446.6 nm, 518.8 nm, 637.8 nm, 659.2 nm, 681.8 nm and 796.2 nm) using two HEXA-BEAM lasers (Photonics Technologies) which allowed for easy wavelength switching without realignment.

Light from the laser passes through a linear polarizer to ensure it is fully polarized before entering the sample. Polarized light then passes through a glass cuvette (Thorlabs CV10 Q3500F) containing an olive oil sample (Tesco Olive Oil) inside a magnetic field, and then through to an analyzer which consists of a second polarizer mounted in a rotation stage (Thorlabs PRM1/MZ8). Finally, the light is incident on a photodiode (Thorlabs PDA100A-EC) which measures its intensity.

The magnetic field was provided by a pair of neodymium (NdFeB) magnets in an aluminium mount; a schematic view of the magnets may be seen in \fref{fig:1} (b). The magnetic field strength could be varied by changing the distance between the magnets. The threaded design of the magnet holders and mount allowed the separation to be adjusted by rotating the magnet holders. Using a hall probe, the magnetic field was measured at various magnet separation distances to determine its spatial homogeneity. Details of the full magnetic field strength calibration may be found in \ref{MagneticfieldModelling}. The same magnets may be used in a wide range of different applications (see for example \cite{So}). 

The rotation stage and photodiode were connected to a PC so that the angle of the analyzer could be controlled and the intensity of light at the photodiode ($I$) could be measured automatically. A computer algorithm was programmed which rotated the analyzer in increments of one degree and then measured the intensity of the light using the photodiode. In this way, a mean intensity with a standard error was measured at each angle over a full 360 degree analyzer rotation. The data were then fitted using Malus' law, as described in (\ref{ModifiedMalus}). The major advantage of using computer-controlled apparatus was the rapid collection of data - a full 360 degree rotation cycle was collected in approximately 10 minutes.

\begin{figure}[t]
\begin{center}
\includegraphics[trim = 0mm 0mm 0mm 0mm, clip, width=12cm]{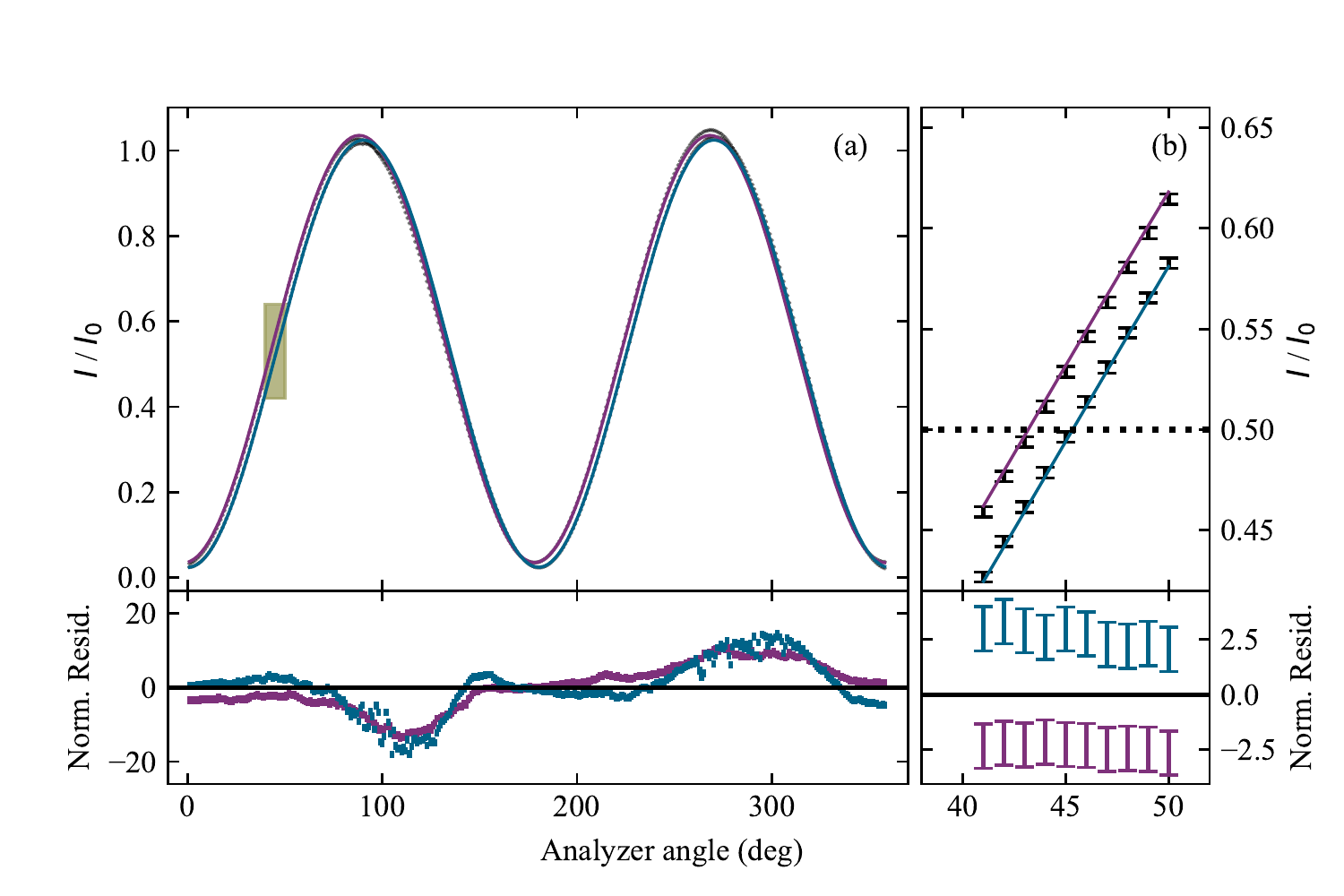}
\caption{The phase shift for a ``combined" sample (olive oil in a glass cuvette) at 446.6 nm. The solid lines are Malus' law models: $I = I_0 \cos^2(\phi + \theta) + c$ which have been normalized by the maximum intensity ($I_0$). The left model (purple) is for data obtained at the minimum magnetic field strength (215.4 $\pm$ 1.4 mT) and the right model (blue) is for data obtained at the maximum magnetic field strength (618.8 $\pm$ 1.4 mT). The lower plots display the normalized residuals for the two different data sets. (a) Transmitted intensity as a function of analyzer rotation. The optical rotation is so small as to be almost imperceptible on  this scale. (b) A zoom of the shaded area in (a) in which the offset $c$ has been subtracted. A dotted line has been added to show the point of half intensity, the measurement point often used in the literature. The residuals show that even after offset subtraction, the data are still fluctuating in this region; hence the need for a rigorous fit to the entire data set (color online).
}
\label{fig:2}
\end{center}

\end{figure}

Applying the Malus' law fit to the data, as demonstrated in \fref{fig:2}, yielded a value of the rotation ($\theta$). Many experiments in the literature will measure rotations by observing the intensity change at a fixed analyzer angle - often 45 degrees where the intensity change will be maximized. However, the residuals of \fref{fig:2} indicate that measurements performed at a single angle are sensitive to changes in the maximum intensity and offset, caused by power fluctuations of the laser and changes in background levels of light. Our method of fitting the model to the full range of the data eliminates this issue by fully parameterizing the amplitude and offset. Hence we may extract rotations precisely, insensitive to intensity fluctuations. 

For each sample being measured, three sets of data were taken at each magnetic field strength to obtain a mean rotation with a standard error. The field strength was then altered and the new value of $\theta$ was calculated using the same procedure. At each magnetic field strength, we compared the measured rotation to the rotation obtained at the minimum field strength (215.4 $\pm$ 1.4 mT). Here we introduce some new notation: $\Delta\theta = \theta - \theta_0$, where $\theta_0$ is the rotation at the lowest field value and $\Delta\theta$ is the change in rotation. Ideally, the rotations would be compared to the sample in the absence of any field but this was not possible as the cuvette had to remain fixed in place during measurements due to the natural birefringence of the glass. However, this would only introduce an offset to the measurements and not affect the gradient from which the Verdet coefficient is determined.

\section{Results}

\begin{figure}[t]
\begin{center}
\includegraphics[trim = 0mm 0mm 0mm 0mm, clip, width=12cm]{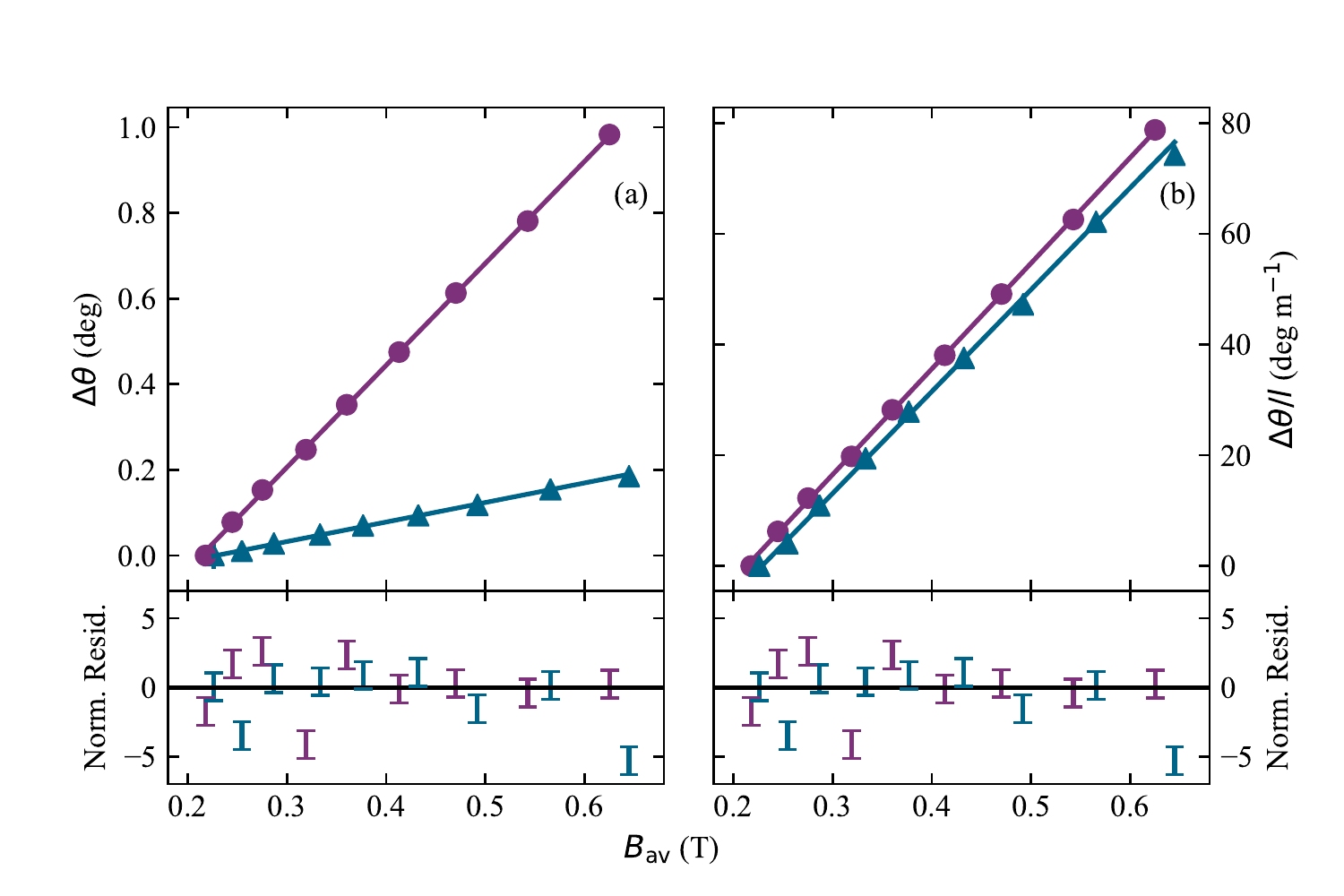}
\caption{Comparison between the empty glass cuvette (triangles) and the ``combined" measurement of the cuvette containing olive oil (circles) for a wavelength of 659.2 nm. (a) The rotation at each value of $B_{\rm av}$ for the two samples. This plot emphasizes that the rotation induced by the glass is large enough that it must be subtracted in order to study the rotation caused by the oil alone. (b) This plot displays the rotations normalized by length such that the gradient is the Verdet coefficient of the material. The Verdet coefficients for glass (see \ref{GlassVerdet}) were found to be very similar to those of olive oil. Thus, $\Delta\theta_{\rm cuvette}$ was only low due to the shorter path length of the glass.  
}
\label{fig:3}
\end{center}

\end{figure}

\begin{figure}[t]
\begin{center}
\includegraphics[trim = 0mm 0mm 0mm 0mm, clip, width=12cm]{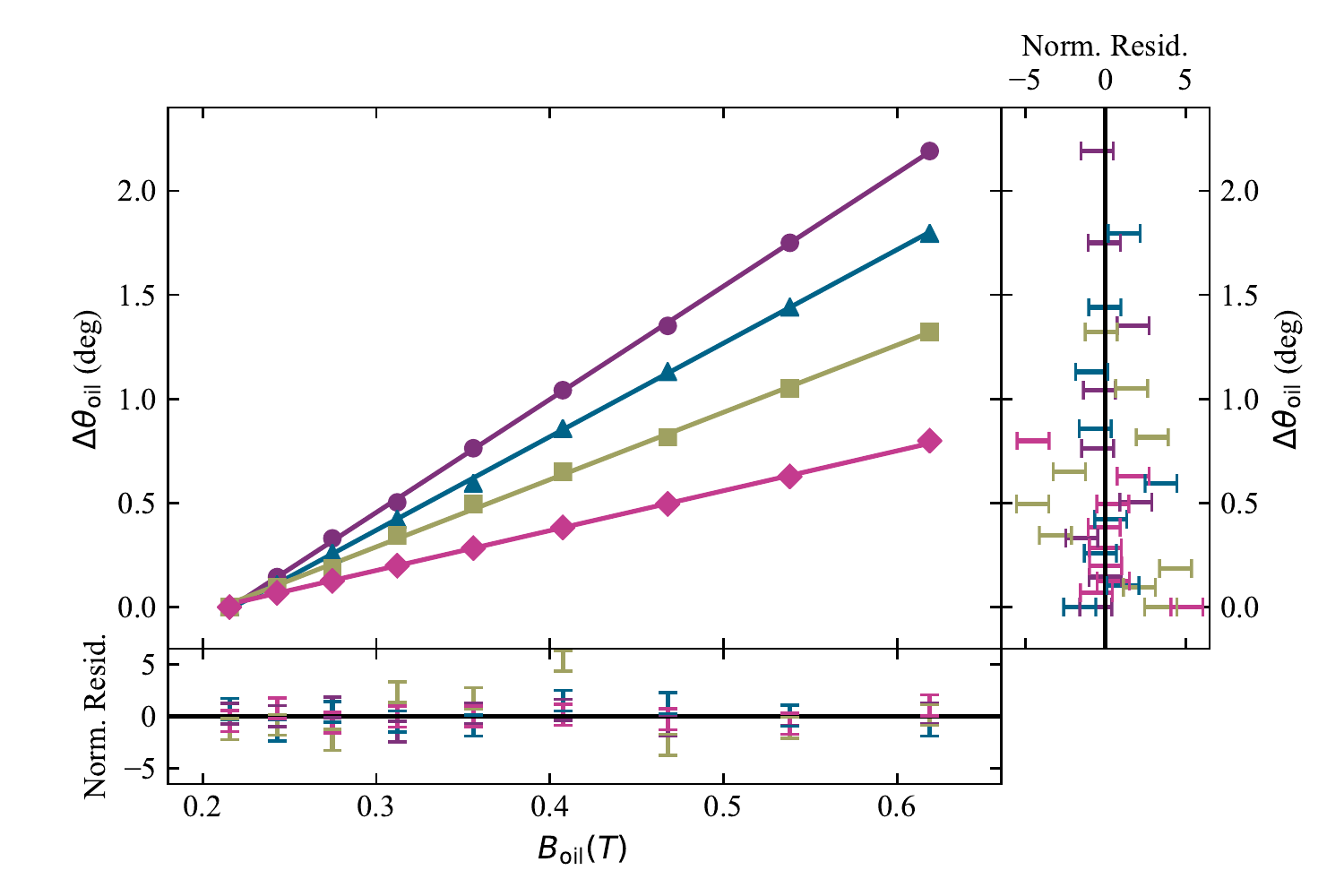}
\caption{The change in rotation due to the oil ($\Delta\theta _{\rm oil}$) for each magnetic field strength at 405.4 nm (circles), 446.6 nm (triangles), 518.8 nm (squares) and 659.2 nm (diamonds). Since $\Delta\theta _{\rm oil} = VB_{\rm oil}l_{\rm oil}$ the data were fit with a straight line using orthogonal distance regression. Error bars are too small to be seen. The lower plot and upper right plot display the normalized residuals in the y direction and x direction respectively. Only 4 of the wavelengths are plotted to avoid too much overlapping of data points.
}
\end{center}
\label{fig:4}
\end{figure}

\begin{figure}[t]
\begin{center}
\includegraphics[trim = 0mm 0mm 0mm 0mm, clip, width=12cm]{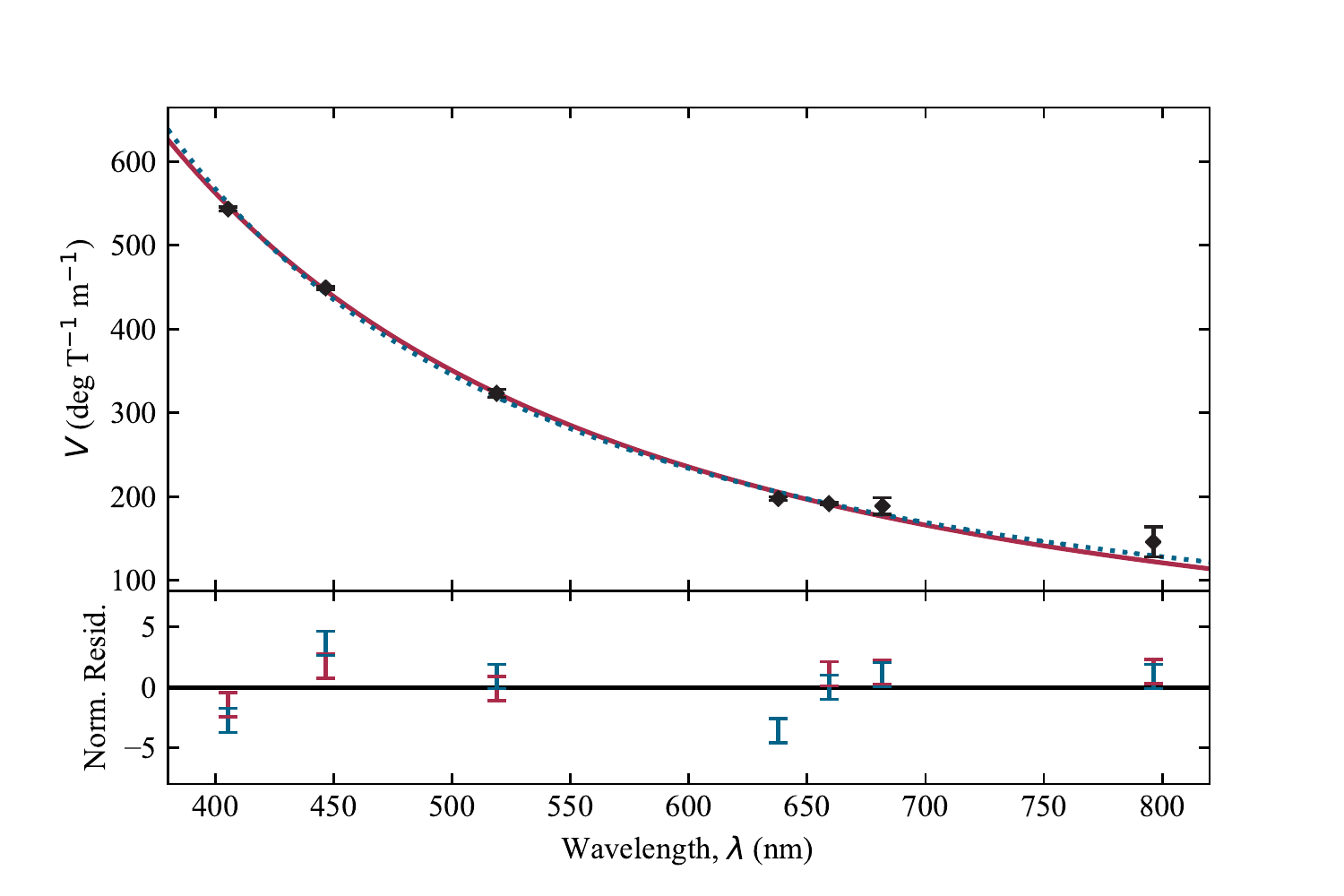}
\caption{Dispersion plot for the variation of the Verdet coefficient of olive oil with wavelength at approximately 20 $^{\circ}$C. Our data are represented by black diamonds with their error bars in the y direction displayed (error bars in x are too small to be seen). The solid red line represents the Cauchy-type model described by $V = A + B/\lambda^2$ and the dotted blue line represents the Drude-type model described by $V = A/(\lambda^2 - \lambda_0^2)$. The $\chi^2_\nu$ values for these models are 4.45 and 7.28 respectively. The lower plot displays the normalized residuals.
}
\label{fig:5}
\end{center}

\end{figure}

\Fref{fig:2} shows example data that have been fit with a weighted least squares algorithm using the Malus' law method according to (\ref{ModifiedMalus}). For the data shown, the rotation change $\Delta\theta $ = 2.19 $\pm$ 0.05 degrees. This is one of the highest values that was observed and emphasizes the need for precise measurements. 

When measuring the Verdet constant of olive oil,  the glass cuvette in which the olive oil is contained also experiences the Faraday effect and so induces extra optical rotation of the transmitted light. This may be seen in \fref{fig:3} where it is clear that the glass makes a significant contribution to the rotation. To account for this, measurements were made of the optical rotation due to the empty glass cuvette. The optical rotation due to the olive oil can then be calculated from a ``combined" measurement of the optical rotation of the cuvette full of oil using \begin{equation}
    \Delta\theta _{\rm oil} = \Delta\theta _{\rm combined} - \Delta\theta _{\rm cuvette},
\label{GlassSubtraction}
\end{equation}
and so the Verdet coefficient may be determined using \begin{equation}
    V = \frac{\Delta\theta _{\rm oil}}{B_{\rm oil}l_{\rm oil}},
\label{SimpleVerdet}
\end{equation}
where $B_{\rm oil}$ is the average field strength experienced by the oil and $l_{\rm oil}$ is the path length of the oil. As this equation implies, a plot of $\Delta\theta_{\rm oil}$ vs. $B_{\rm oil}$ allows the Verdet coefficient to be calculated by dividing the gradient by the path length.

As described in section \ref{Methods}, the Malus' law fitting procedure was repeated to obtain $\Delta\theta _{\rm oil}$ values at each field strength such that a linear fit could be performed. The linear fits for 4 of the wavelengths are displayed in \fref{fig:4}. Because there were errors in $B_{\rm oil}$ values and in $\Delta\theta _{\rm oil}$, the data were fit using an orthogonal distance regression algorithm. The errors on the fitting parameters were obtained from the square roots of the relevant entries in the covariance matrix (procedure as outlined in \cite{Measurements}). The uncertainty in the Verdet constant was then propagated from the error in the gradient and the error in the path length. Table \ref{Table 1} gives the measured Verdet coefficients and uncertainties at each wavelength.

As shown in \fref{fig:5}, the measured Verdet coefficients were fit with Cauchy-type and Drude-type dispersion curves using (\ref{Cauchy}) and (\ref{Drude}) respectively. For the Cauchy-type fit, the parameters were found to be $A$ = -26 $\pm$ 4 deg T$^{-1}$ m$^{-1}$ and $B$ = 9.42 $\pm$ 0.12 $\times$ $10^7$ deg T$^{-1}$ m$^{-1}$ nm$^2$ while for the Drude-type fit they were $A$ = 7.9 $\pm$ 0.2 $\times$ 10$^{7}$ deg T$^{-1}$ m$^{-1}$ nm$^2$ and $\lambda_0$ = 142 $\pm$ 13 nm. As before, the errors on these fitting parameters were obtained from the square root of the relevant entries in the covariance matrix. 

\begin{table}
\caption{\label{Table 1} The determined Verdet coefficients for olive oil at approximately 20 $^{\circ}$C.}
\begin{indented}
\item[]\begin{tabular}{@{}cc}
\br 
    Wavelength $\pm$ 0.1 (nm) & Verdet coefficient (deg T$^{-1}$ m$^{-1}$) \\
\mr
    405.4 & 543 $\pm$ 2\\
    446.6 & 449 $\pm$ 2\\
    518.8 & 323 $\pm$ 5\\
    637.8 & 198 $\pm$ 2\\
    659.2 & 192 $\pm$ 1\\
    681.8 & \lineup{\0}190 $\pm$ 10\\
    796.2 & \lineup{\0}146 $\pm$ 18\\
\br
\end{tabular}
\end{indented}
\end{table}

\section{Discussion} \label{Discussion}

The dispersion relationship we determined is similar to that of other liquids such as water \cite{Jain}. Our determined Verdet coefficients are also in good agreement with recent values present in the literature at similar wavelengths \cite{Brandon}. Similar to these recent measurements, we did not observe an anomalously high Verdet coefficient at 650 nm as was previously reported \cite{Shakir}. 

A restriction of using our permanent magnet set up is that much smaller path lengths must be used than, for example, in a solenoid set up. However, the high field strengths we were able to achieve help compensate for this and so measurable rotations are still easily observed. The path length of the oil was measured precisely using digital calipers so despite being small it did not introduce major uncertainty into the Verdet coefficient values. 

We believe the major strength of our method is the rigorous Malus' law fitting which allows us to average over fluctuations in the laser power and detect rotations with errors on the order of tens of $\mu$rads. Performing the fit and allowing for an offset to be paramaterized makes the method insensitive to changes in the background level of light which is often difficult to control, particularly in a teaching laboratory. 

Traditionally, measuring the intensity over a full range of angles would be a tedious process, requiring manual rotation of the analyzer. Using the computer-controlled photodiode and analyzer makes the data collection much easier and faster. This means a Verdet coefficient may be determined even in a short laboratory session. The coding of the algorithm is also a vital skill as in modern physics laboratories data collection is becoming increasingly automated. The coding can either be partially or fully completed by students to give them an understanding of how the computer is interfacing with the equipment.

As shown in \fref{fig:2}, the raw data give an excellent visual demonstration of Malus' law which students should be familiar with from earlier levels of teaching. Performing the fit provides students with a good opportunity to develop their model fitting and error analysis skills. 

Depending on the level of students, many aspects of the experiment may be partially or fully completed for them, to tailor the learning to their skill set. For instance, if they are not expected to be able to perform non-linear fits, they could be provided with the Malus' law fitting algorithm. In particular, the full process of calibrating the magnetic field, as set out in \ref{MagneticfieldModelling}, was fairly advanced and for students it should be sufficient to determine $B_{\rm oil}$ at each magnet separation simply by taking an average of several measurements. The full calibration then presents an interesting area in which students can extend the experiment. Other avenues for extension include attempting to use the Verdet coefficient to detect adulteration of olive oil (similar to another experiment in the literature \cite{Adulteration}) or measuring the Verdet coefficient of other fluids. 

\section{Conclusions}

We have devised a simple experimental procedure based on Malus' law which measures polarization rotations with a precision of up to 50 $\mu$rads. The versatility of the Malus' law fitting method allows it to be used in a variety of different polarimetry experiments including the study of chirality in sugars, stress-induced birefringence in plastics and magneto-induced birefringence in liquids and gases. To prove the efficacy of our method, we studied the Faraday effect in olive oil and were able to measure Verdet coefficients with errors as small as 0.4\%. Our results agree well with values in the literature and we present excellent fits to dispersive models. The experiment is simple in its execution and equipment, yet can provide profound insight into fundamental concepts in optics and electromagnetism and as such is well-suited to being used as an undergraduate teaching experiment. The experiment provides training in a wide range of experimental skills, including data analysis and error analysis. In addition, the versatility of the apparatus provides ample opportunity for extension.

\section*{Acknowledgements}
We would like to thank Dr Aidan Hindmarch and Dr Jason Anderson for providing data aquisition software, Chloe So for help and advice, and Durham University for providing the equipment and funding required to complete this investigation.

NLRS and CSA acknowledge financial support from EPSRC Grant Ref. No. EP/M014398/1. IGH and CSA acknowledge financial support from EPSRC Grant Ref. No. EP/R002061/1.

\section*{References}



\appendix

\section{Magnetic field modelling} \label{MagneticfieldModelling}

The magnetic field was calibrated to determine the spatial homogeneity and average field strength at each magnet separation distance. Each full rotation of the magnets in their mount increased their separation by approximately 2 mm and therefore reduced the magnetic field.

\begin{figure}[h]
\begin{center}
\includegraphics[trim = 0mm 0mm 0mm 0mm, clip, width=12cm]{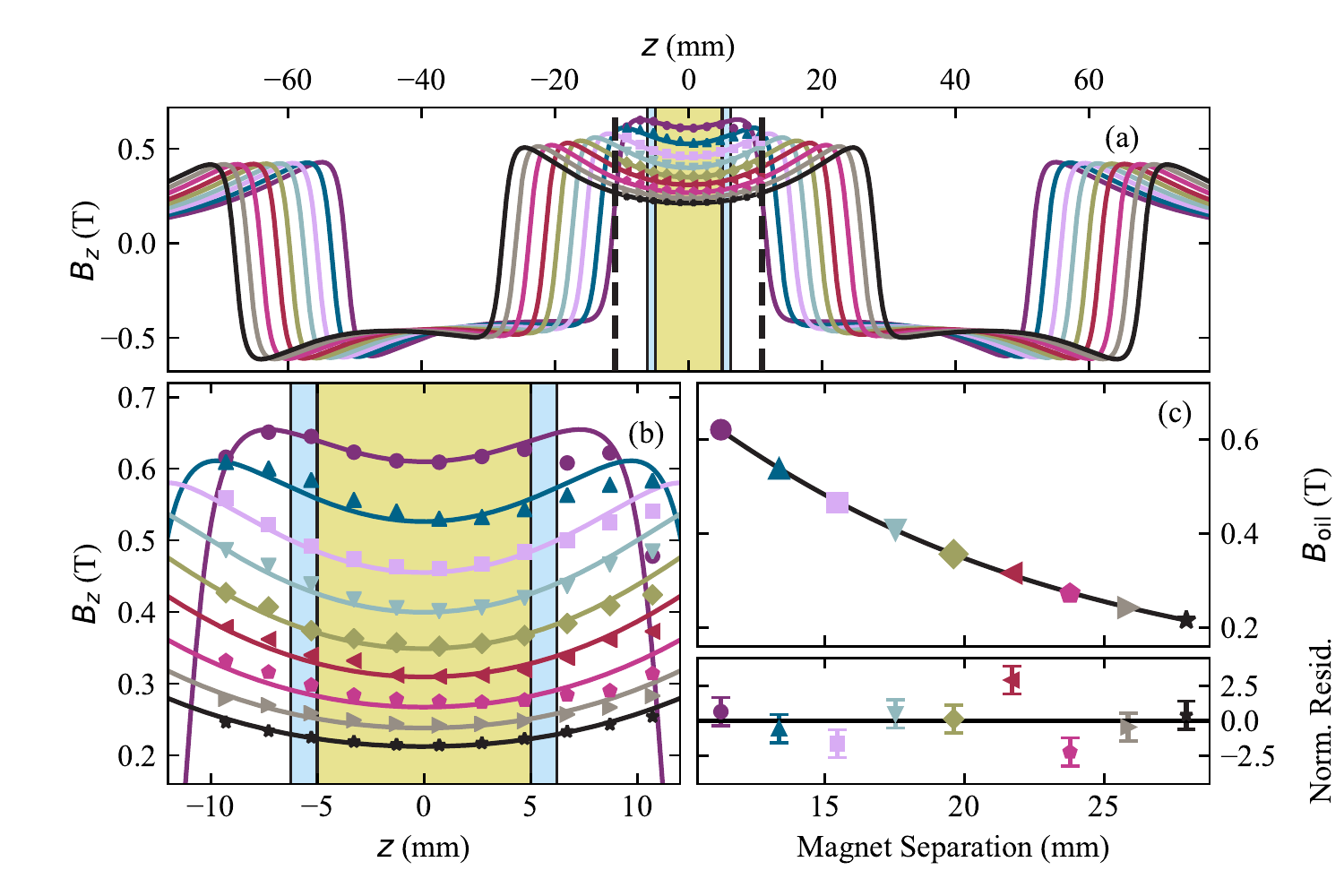}
\caption{(a) The spatial homogeneity of the magnetic field along the \textit{z} direction for each separation distance. The point \textit{z} = 0 represents the position equidistant from the magnets, at which the centre of the sample is placed. The theory curves are fitted according to (\ref{AnalyticField}) The dotted lines show the minimum separation of the magnets i.e. the spatial region in which data were collected. (b) A zoomed-in plot of the region between the dotted lines of the upper figure. This plot shows the effect of altering the magnet separation on the field strength in the region between the magnets. The outer shaded region shows the position of the glass walls of the cuvette ($l_{\rm glass}$ = 2.48 $\pm$ 0.01 mm) and the inner shaded region shows the position of the oil ($l_{\rm oil}$ = 10.00 $\pm$ 0.01 mm).  (c) The variation of the average magnetic field strength in the oil (\textit{$B_{\rm oil}$}) as a function of the magnet separation. The lower plot displays the normalized residuals. (color online).
}
\label{fig:A1}
\end{center}

\end{figure}

The spatial homogeneity was measured using a transverse hall probe which was moved along the direction of laser propagation through the magnetic field (\textit{z} axis). These measurements are summarized in  \fref{fig:A1}(a) and (b). Theory curves were fit to the data by considering the axial field strength $B_z$ at a distance $z$ from a cylindrical magnet with uniform magnetization $B_0$ and radius $R$: 
\begin{equation}
    B_z = \frac{B_0}{2} \left[\frac{z + z_0 + t}{\sqrt{(z + z_0 + t)^2 + R^2}} - \frac{z + z_0 - t}{\sqrt{(z + z_0 - t)^2 + R^2}}\right]
\label{AnalyticField}
\end{equation}
where $z_0$ is an offset to account for the fact that the magnet is not at the origin and $t$ is the thickness of the magnet (for more detail see \cite{Weller:thesis}). As seen in \fref{fig:1}(b), the magnets were not perfect cylinders and so the field produced by the magnet was determined by piecewise addition of the field due to each part of the magnet and a subtraction due to the hole bored through the magnet's centre. It was then possible to sum the contributions from each magnet to work out the overall field strength at each $z$ position.

The theory curves were used to calculate the average field strength experienced by the oil ($B_{\rm oil}$) for each separation distance; these results are summarized in \fref{fig:A1}(c). It should be noted that the average field strength experienced by the combined sample ($B_{\rm combined}$) and cuvette ($B_{\rm cuvette}$) are not the same as for the oil alone. However, as shown by (\ref{SimpleVerdet}), $B_{\rm oil}$ is the only field strength required to calculate the Verdet coefficient of the oil.

\section{Verdet coefficients in glass}
\label{GlassVerdet}

As discussed in the results section, to determine the Verdet coefficient for olive oil it was necessary to perform a separate set of measurements with an empty cuvette to subtract the rotation induced by the glass. A side-effect of performing these measurements is that we were able to determine Verdet coefficients for the UV fused quartz glass at each of the wavelengths we tested; these are summarized in table \ref{Table B1}. In general, the glass Verdet coefficients have a larger uncertainty than those for the oil, this could be due to the shorter path length of the glass.  

\begin{table}
\caption{\label{Table B1} The determined Verdet coefficients for the UV fused quartz glass cuvette at approximately 20 $^{\circ}$C.}
\begin{indented}
\item[]\begin{tabular}{@{}cc}
\br 
    Wavelength $\pm$ 0.1 (nm) & Verdet coefficient (deg T$^{-1}$ m$^{-1}$) \\
\mr
    405.4 & \lineup{\0}540 $\pm$ 16\\
    446.6 & 431 $\pm$ 5 \\
    518.8 & \lineup{\0}293 $\pm$ 13 \\
    637.8 & \lineup{\0}196  $\pm$ 11 \\
    659.2 & 181 $\pm$ 2\\
    681.8 & \lineup{\0}190 $\pm$ 30\\
    796.2 & \lineup{\0}370 $\pm$ 50 \\
\br
\end{tabular}
\end{indented}
\end{table}

\begin{figure}[t]
\begin{center}
\includegraphics[trim = 0mm 0mm 0mm 0mm, clip, width=12cm]{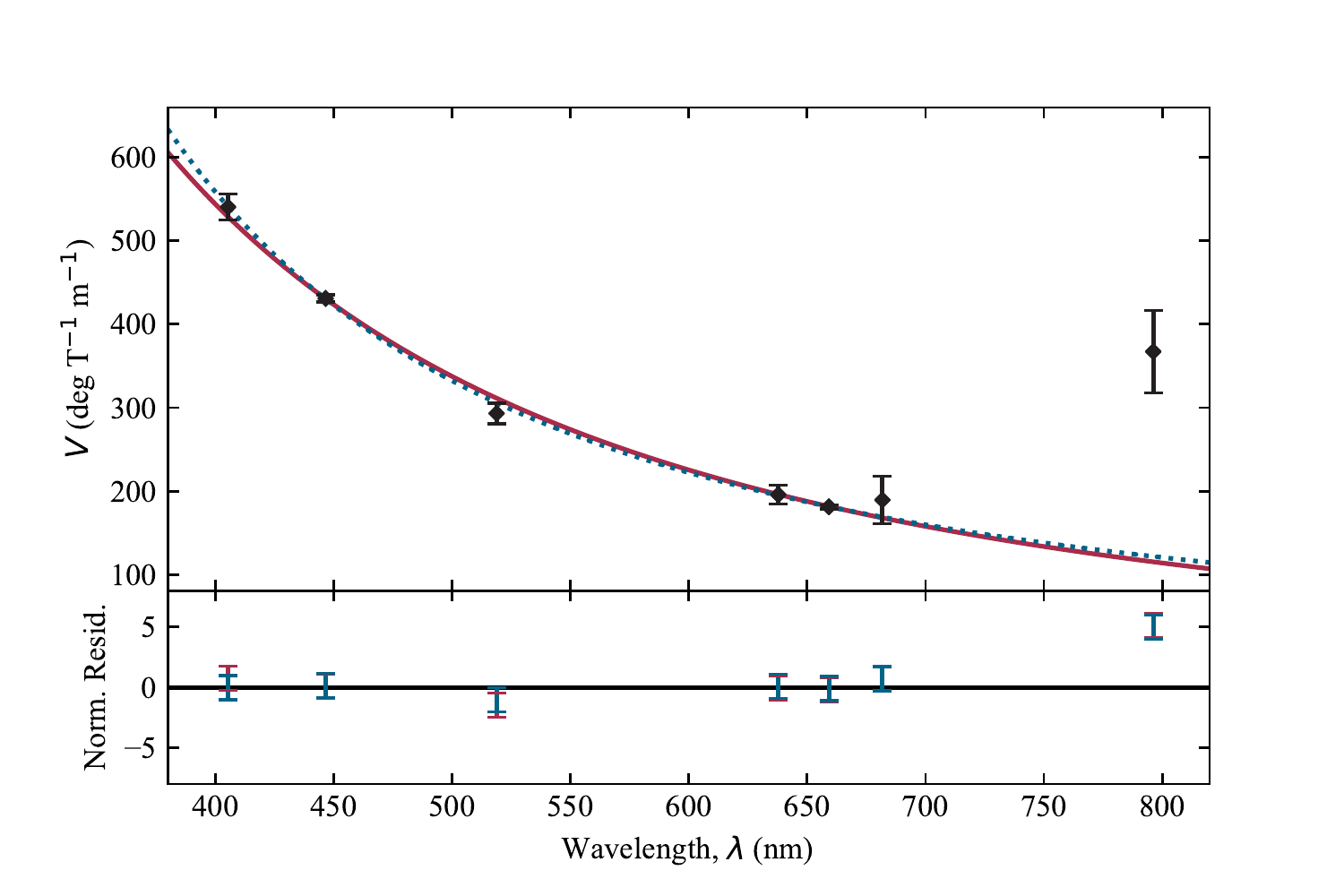}
\caption{Dispersion plot for the variation of the Verdet coefficient of quartz glass with wavelength at approximately 20 $^{\circ}$C. The data are represented by black diamonds with their error bars in the y direction displayed (error bars in x are too small to be seen). The solid red line represents the Cauchy-type model described by $V = A + B/\lambda^2$ and the dotted blue line represents the Drude-type model described by $V = A/(\lambda^2 - \lambda_0^2)$. The $\chi^2_\nu$ values for these models are 5.86 and 5.26 respectively. The lower plot displays the normalized residuals. The point corresponding to 796.2 nm does not fit the model well and has a very large error; this could be due to incomplete extinction of the polarizers or a separate absorption peak in the IR region.
}
\label{fig:B1}
\end{center}

\end{figure}

\Fref{fig:B1} shows the dispersion curves applied to our data for the glass. As with the olive oil data, the dispersive models fit well and our data is in good agreement with the literature \cite{Ramaseshan}. The exception is the data point corresponding to 796.2 nm which is far from the models and has a large error. We believe this may be due to the polarizers being unsuited to the near-IR region. Alternatively, the quartz glass may possess an absorption peak in the IR region, such that this wavelength is actually part of a separate dispersion curve - though this would require further study to confirm.  Whilst we focused on the Faraday effect in olive oil, we show here that our method may be used more generally and also works well for the Faraday effect in solids.

\end{document}